\documentclass[longauth]{aa}  
\usepackage{graphicx}
\usepackage{subfig}
\usepackage{natbib}
\usepackage{color}
\usepackage{amsmath}
\usepackage[normalem]{ulem}

\newcommand{\asec}{$^{\prime\prime}$}
\definecolor{gris}{gray}{0.5}

\begin{document}

   \title{Tracing extended low-velocity shocks through SiO emission
   }

   \subtitle{Case study of the W43-MM1 ridge}

   \author{F. Louvet\inst{1,2}
          \and F. Motte\inst{1,3}
          \and A. Gusdorf\inst{4}
          \and Q.~Nguy$\tilde{\hat{\rm e}}$n~Lu{\hskip-0.65mm\small'{}\hskip-0.5mm}o{\hskip-0.65mm\small'{}\hskip-0.5mm}ng\inst{5}
          \and P. Lesaffre\inst{4}
          \and A. Duarte-Cabral\inst{6}
          \and A. Maury\inst{1}
          \and N. Schneider\inst{7}
          \and T. Hill\inst{8}
          \and P. Schilke\inst{7}     
          \and F. Gueth\inst{9}
          }

\institute{Laboratoire AIM Paris-Saclay, CEA/IRFU - CNRS/INSU - Universit\'e Paris Diderot, Service d'Astrophysique, B\^at. 709, CEA-Saclay, F-91191 Gif-sur-Yvette Cedex, France, 
              \email{fabien.louvet@cea.fr}
\and Departamento de Astronomia de Chile, Universidad de Chile, Santiago, Chile
\and Institut de Planétologie et d'Astrophysique de Grenoble (IPAG), France
\and LERMA, UMR 8112 du CNRS, Observatoire de Paris, \'Ecole Normale Sup\'erieure, 24 rue Lhomond, 75231 Paris Cedex 05, France           
\and ALMA Chile Observatory, National Observatory of Japan 
\and School of Physics and Astronomy, University of Exeter, Stocker Road, Exeter EX4 4QL, UK
\and  Physikalisches Institut, Universit\"at zu K\"oln, Z\"ulpicher Stra\ss{}e 77, 50937 K\"oln, Germany
\and Joint ALMA observatory, Santiago, Chile
\and  Institut de Radioastronomie Millim\'etrique (IRAM), 300 rue de la Piscine, 38406 Saint Martin d'H\`eres, France
%
        }
     \date{Received 2016; accepted 2016}

 
  \abstract
   {}   
   {Previous literature suggests that the densest structures in the interstellar medium form through colliding flows, but patent evidence of this process is still missing. Recent literature proposes using SiO line emission to trace low-velocity shocks associated with cloud formation through collision. In this paper we investigate the bright and extended SiO(2--1) emission observed along the $\sim$5\,pc-long W43-MM1 ridge to determine its origin.}
   {We used high angular resolution images of the SiO(2--1) and HCN(1--0) emission  lines obtained with the IRAM plateau de Bure (PdBI) interferometer and combined with data from the IRAM 30\,m radiotelescope. These data were complemented by a \textit{Herschel} column density map of the region. We performed spectral analysis of SiO and HCN emission line profiles to identify protostellar outflows and spatially disentangle two velocity components associated with low- and high-velocity shocks. Then, we compared the low-velocity shock component to a dedicated grid of one-dimensional (1D) radiative shock models.
}
   {We find that the SiO emission originates from a mixture of high-velocity shocks caused by bipolar outflows and low-velocity shocks. Using SiO and HCN emission lines, we extract seven bipolar outflows associated with massive dense cores previously identified within the W43-MM1 mini-starburst cluster. Comparing observations with dedicated Paris-Durham shock models constrains the velocity of the low-velocity shock component from 7 to 12\,km.s$^{-1}$. }   
   {The SiO arising from low-velocity shocks spreads along the complete length of the ridge. Its contribution represents at least 45\% and up to 100\% of the total SiO emission depending on the area considered. The low-velocity component of SiO is most likely associated with the ridge formation through colliding flows or cloud-cloud collision.}

\titlerunning{SiO emission and 
colliding flows
in W43-MM1}
\authorrunning{Louvet et al.}
 \maketitle
%

\section{Introduction}
High-mass stars (OB-type, $\geq$8\,M$_\odot$) are known to form in massive dense cores \citep[MDCs, diameter of $\sim$\,0.1\,pc and density $>$\,10$^5$\,cm$^{-3}$; see, e.g.,][]{motte07}. But the formation of MDCs, as well as the physical process by which high-mass stars form within MDCs, are still badly constrained. The formation of high-mass stars is explained in two ways: quasi-static and dynamical scenarios. The quasi-static scenario describes a monolithic collapse of the MDC that is supported by supersonic turbulent pressure \citep[e.g.,][]{mckee2002}. However, the dynamical scenarios suggest either a coalescence of low/intermediate-mass protostars \citep[e.g.,][]{bonnell-bate02} or a formation of high-mass stars through the interplay of colliding flows associated with the cloud formation \citep[e.g.,][]{bonnell-bate02, heitsch08}. In this latest dynamical picture, cloud formation generates colliding flows that funnel mass from large potentials to small scales \citep[e.g.,][]{vazquezsemadeni05, hartmann12, smith13}. While the gas flows are colliding, we expect low-velocity shocks at the colliding interfaces. 

At the MDC scale ($\sim$0.1\,pc) dynamical signatures, such as gravitational streams and shearing motions, have been found using N$_2$H$^+$ and H$^{13}$CO$^+$ lines \citep[see, e.g.,][]{csengeri11a,csengeri11b,henshaw14}. At that scale, low-velocity shocks have also been reported using CH$_3$CN and SiO lines \citep[e.g.,][]{csengeri11b, duarte-cabral14}.

The \textit{Herschel} key program HOBYS \cite[see][]{motte10,motte12} demonstrated that MDCs, and later high-mass stars, tend to form in high-density filaments with typical sizes of $\sim$1--10\,pc, above 10$^{23}$\,cm$^{-2}$ in column density. These structures are called ridges \citep{hill11, quang11-w43, martin12}.

On the scale of ridges, gas inflow, global collapse, and velocity gradients have been reported by several groups \citep[e.g.,][]{schneider10,quang13,peretto13,tackenberg14}. However, all the studies suggesting that SiO emissions could trace low-velocity shocks associated with the collision of gas inflows \citep{jimenez10, quang11-w43, quang13, sanhueza13} were impaired by their limited sensitivity and/or angular resolution. All of these studies reported the SiO emission line profile as the association of two Gaussian components at approximately the same central velocity. While the broad Gaussian component was attributed to high-velocity shock linked to protostellar outflows, the narrow Gaussian component was equally attributable to low-velocity shock of either cloud-cloud collision or less powerful outflows beyond the angular resolution performances.


\begin{table*}[htbp!]
\caption{Main observational parameters}
\label{t:pariram}
\begin{center}
\begin{tabular}{c|cc|cc}
\hline
\hline
Parameter                              & \multicolumn{2}{c}{IRAM PdBI}                                               & \multicolumn{2}{c}{IRAM 30\,m}                         \\
                                       & SiO (2-1)                            & HCN (1-0)                            & SiO (2-1)                     & HCN (1-0)            \\
\hline                                                                                                                                                                        
Frequency (GHz)                        & 86.846                               & 88.631                               & 86.846                        & 88.631               \\
Bandwidth (MHz)                        & 40                                   & 80                                   & --                            & --                   \\
Spec. Res. (km.s$^{-1}$)               & 0.27                                 & 1.06                                 & 0.67                          & 0.67                 \\
Primary beam                           & 59\arcsec                            & 59\arcsec                            & 29.9\arcsec                   & 29.9\arcsec          \\                                                        
Synthesized beam                       & $4\farcs96\times3\farcs14$           & $5\farcs02\times3\farcs08$           & -                             & -                    \\
$1\sigma$ rms                          & 0.20 Jy/Beam.km/s                    & 0.49 Jy/Beam.km/s                    & 0.02 K.km/s                   & 0.04 K.km/s          \\
System temperature (K)                 & 120                                  & 120                                  & -                             & -                      \\   
\hline
\end{tabular}
\end{center}
\end{table*}


We were able to disentangle these broad and narrow Gaussian components in the most extreme of the ridges: \object{W43-MM1}. This ridge lies in the W43 molecular complex \citep{quang11-w43,carlhoff13, motte14}. A distance from the sun of $5.5\pm0.4$\,kpc was inferred for this complex by parallax measurements \citep{zhang14}. Given their uncertainties, and since the measurements were based on only four sources located at the periphery of the complex, we adopted a round 6\,kpc distance, which is consistent with \cite{louvet14}.
\cite{louvet14} modeled \object{W43-MM1} using a 3.9~pc$\times$2~pc$\times$2~pc ellipsoid with a total mass of $\sim$2$\times 10^4$~M$_\odot$ and an average density of $\sim$4.3$\times 10^4$ cm$^{-3}$. This ridge hosts a protocluster of 12 MDCs of mean size of $\sim$0.07\,pc and masses ranging from 20\,M$_\odot$ to 2000\,M$_\odot$ \citep{louvet14,sridharan14}. It undergoes a remarkable burst of (high-mass) star formation with an instantaneous star formation rate of $\sim$6000\,M$_{\odot}$.Myr$^{-1}$ \citep{louvet14}. 

As part of the W43-HERO IRAM large program, \cite{quang13} discovered a bright and extended SiO(2--1) emission with $N_{\rm SiO}\sim 6\times10^{13}$~cm$^{-2}$ over a $\sim$43~pc$^2$ area along and around \object{W43-MM1}. They interpreted this SiO emission as arising from a low-velocity cloud collision the ridge experienced during its formation, but their observations lacked the resolution to rule out an origin from a protocluster. 
 
In this paper, we show that SiO emission can be an excellent tracer of colliding flows. Making use of the interferometric and single-dish data cubes described in Sect.~\ref{s:obs}, we investigate the high-density parts of \object{W43-MM1}, look for outflows driven by protostars forming in the \object{W43-MM1} ridge, and quantify the SiO intensity unambiguously associated with low-velocity shocks in Sect.~\ref{s:analysis}. Section~\ref{s:model} confronts the integrated SiO emission values with dedicated shock models. Finally, Sect.~\ref{s:conclusion} gives our conclusions.

\section{Observations and reduction}
\label{s:obs}

\subsection{IRAM PdBI mosaic}

A seven-field 3~mm mosaic of the \object{W43-MM1} ridge has been carried out with the IRAM Plateau de Bure Interferometer (PdBI). Configurations  C \& D  (baseline lengths from 24 to 229 m)  were used in October \& July 2011 with 5 and 6 antennas respectively. Broadband continuum shown in \cite{louvet14} and spectral lines were simultaneously observed. The phase, amplitude, and correlator bandpass were calibrated on strong quasars (3C454.3, 1827+062, and 0215+015), while the absolute flux density scale was derived from MWC349 observations. The average system temperature was of $\sim$120~K and absolute flux calibration uncertainty was estimated to be $\sim$15\%. 

Three spectral units of 40 MHz bandwidth with individual channel spacing of 78 kHz were tuned to observe the SiO(2--1) line at 86.846~GHz. A second spectral unit of the correlator was tuned to the HCN(1--0) line at 88.983 GHz with a total bandwidth of 80 MHz and individual channel spacing of 313 kHz. The observational parameters of the IRAM PdBI data are summarized in Table~\ref{t:pariram}.

The noise level of the SiO(2--1) and HCN(1--0) data cubes was affected by the continuum emission, which was especially high in the densest part of the \object{W43-MM1} ridge. To correct for this artifact, we subtracted the continuum emission built from the WIDEX backend from each line, and fully described in \cite{louvet14}. This process improved the mean sensitivity of integrated maps by $\sim$20\%. We used the GILDAS\footnote{See the following web page for details: http://www.iram.fr/IRAMFR/GILDAS/} package to calibrate each field data set, merged the visibility data of the seven mosaic fields, then inverted and cleaned (natural cleaning) the resulting IRAM PdBI image.

\begin{figure*}[htbp!]
\begin{center}
\includegraphics[trim=0cm 0cm 0cm 0cm, clip, scale=0.85]{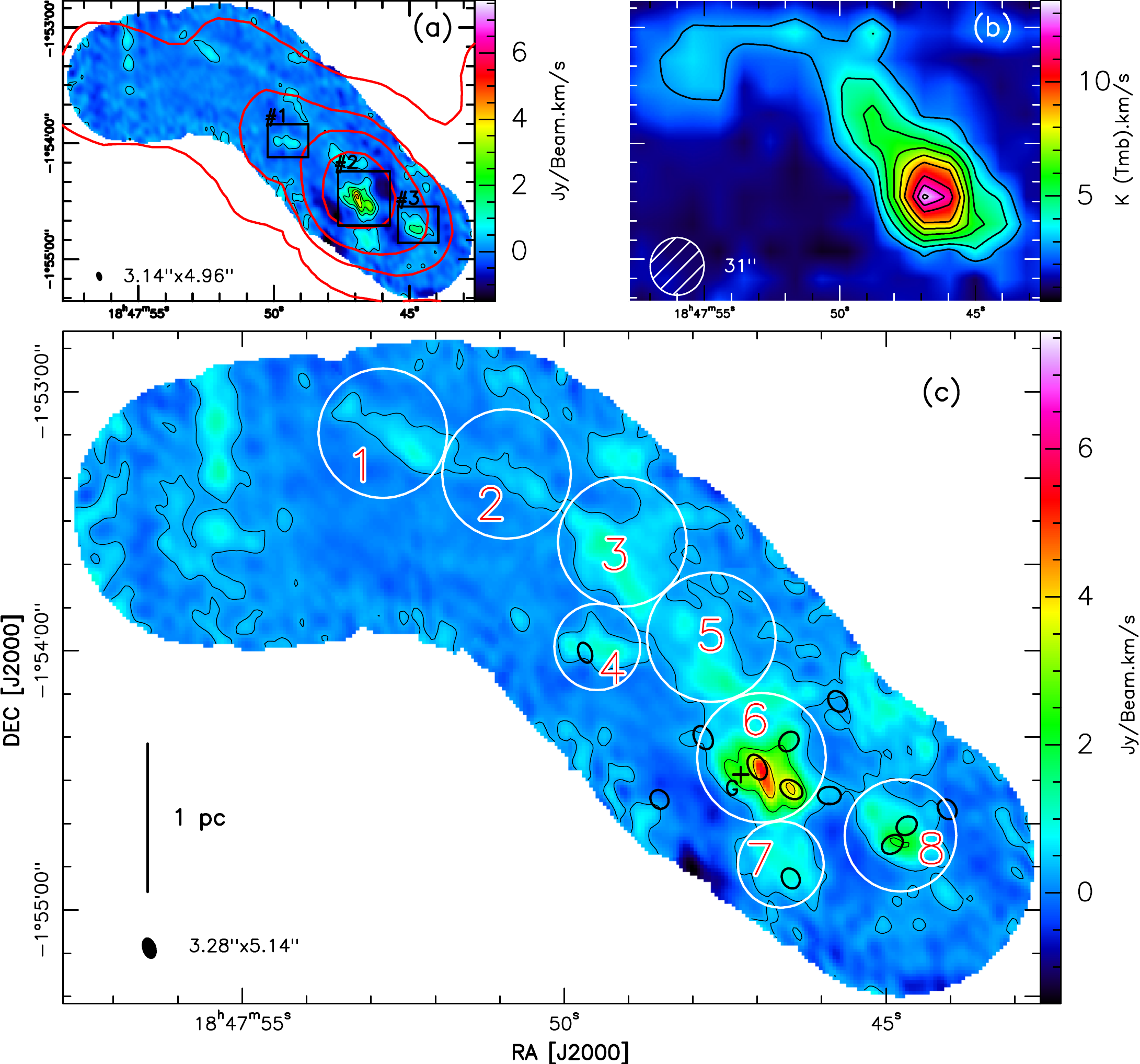}
\caption{Extended SiO emission toward the \object{W43-MM1} ridge. SiO(2--1) maps, integrated from 80 km.s$^{-1}$ to 120 km.s$^{-1}$, were obtained with the IRAM PdBI (a), IRAM 30 m (b), and from the merging of both data sets (c). \textbf{Panel (a):} black contours start at 1.9\,Jy/beam.km/s with 0.95\,Jy/beam.km/s steps. Red contours correspond to the \textit{Herschel} column density map \citep[see][]{quang13, louvet14}. The contours, going towards the inner parts, are $5\times10^{22}$\,cm$^{-2}$, $10^{23}$\,cm$^{-2}$, $1.8\times10^{23}$\,cm$^{-2}$, and $3.5\times10^{23}$\,cm$^{-2}$. Black rectangles, labeled \#1, \#2, and \#3, indicate the areas where zooms are shown in Fig.~\ref{f:flot}. \textbf{Panel (b):} from \citet{quang11-w43}; black contours start at 2.8\,K.km/s with 1.2\,K.km/s steps. \textbf{Panel (c):} the black contours start at 0.3\,Jy/beam.km/s with 1.8\,Jy/beam.km/s steps. Black ellipses highlight MDCs extracted from the interferometric continuum maps presented in \cite{louvet14} and the black cross indicates an extra source, named G, presented by \citet{sridharan14}. White circles numbered from 1 to 8 indicate the positions of the integrated SiO(2-1) emission spectra presented in Fig.~\ref{f:raies}.}
\label{f:compa}
\end{center}
\end{figure*}

As an example, Fig.~\ref{f:compa}a shows the IRAM PdBI SiO(2--1) map integrated from 80~km.s$^{-1}$ to 120~km.s$^{-1}$.bThe effective spatial resolutions of the IRAM PdBI maps are $\sim$5\asec$\times$3\asec, corresponding to $\sim$0.11~pc at the distance of 6~kpc used for \object{W43-MM1} (see Table~\ref{t:pariram}).

\subsection{Short spacing from the IRAM 30\,m}

We extracted the SiO(2--1), HCN(1--0), emission lines from a $\sim$300 arcmin$^{2}$ mapping survey of the W43-Main mini-starburst region performed at 3\,mm (85--93 GHz). These observations were carried out as part of the IRAM 30 m large program W43-HERO\footnote{The W43-HERO (W43 Hera/EmiR Observation) project is an IRAM 30~m large program, whose description and data can be accessed at: http://www.iram-institute.org/EN/content-page-292-7-158-240-292-0.html} dedicated to the origins of molecular clouds and star formation in W43 (\citealt{quang13} ; \citealt{carlhoff13}).

The program was observed with the Eight MIxer Receiver (EMIR) in December 2010, January 2011, and March 2011. It used low spectral resolution ($\sim$0.67 km.s$^{-1}$) but large bandwidth (8 GHz) fast Fourier transform spectrometer (FTS) backend. 
The map was completed using the on-the-fly mode in two perpendicular scanning directions (R.A. and Dec.). The pointing error was less than 3\arcsec ~and calibration accuracy was within 10\%. 

To remove the baseline in each spectrum, we masked the 80--120~km.s$^{-1}$ velocity range of the broadest integrated line. First-order polynomials were used  to fit the masked spectra, derive the baseline, and remove the baseline from the entire spectra. We then combined the reduced spectra into a sampled data cube with a 10\arcsec~Gaussian kernel, providing a $\sim$30\asec~angular resolution.
The observational parameters of the SiO(2--1) and HCN(1--0) data are summarized in Table~\ref{t:pariram}. 
Finally, we converted the units of the data cubes from antenna temperature, $T_{\rm A^*}$, to main beam temperature, $T_{\rm mb}$, following the usual conversion, $T_{\rm mb}$ = F$_{\rm eff}$/B$_{\rm eff} \times$T$_{A^*}$ in K with F$_{\rm eff}$/B$_{\rm eff}$ = 1.23\footnote{The efficiency is provided on the IRAM website: http://www.iram.es/IRAMES/mainWiki/Iram30mEfficiencies.} at 87~GHz.

Figure~\ref{f:compa}b shows the IRAM 30 m SiO(2--1) map integrated from 80~km.s$^{-1}$ to 120~km.s$^{-1}$. The figure illustrates the $\sim$5\,pc~$\times$1.5~pc extent of the large-scale SiO emission published by \cite{quang13} and targeted by the IRAM PdBI mosaic.

\subsection{Combining IRAM PdBI and IRAM 30 m data}
\label{s:reduction}

The IRAM PdBI offers better angular and spectral resolutions and better line sensitivities than the IRAM 30 m data (compare Fig.~\ref{f:compa}a and Fig.~\ref{f:compa}b; see Table~\ref{t:pariram}). The IRAM/PdBI mosaic however filters out the extended emission (filtering scale $\sim$18\arcsec), which is crucial to retrieve the emission arising from the bulk of the cloud. Moreover, the cleaning process is less efficient and leads to artefacts when short spacings are missing \citep[e.g.,][]{gueth-guilloteau00}. Therefore, we merged the IRAM PdBI mosaic and IRAM 30 m image. The main idea was to take advantage of the large diameter of the IRAM 30 m antenna to fill the central hole of the IRAM PdBI uv-coverage. The algorithm used to merge IRAM 30 m and IRAM PdBI data extracted pseudo-visibilities from the single dish map for each pointing center of the mosaic. The interferometer and single-dish visibilities were then processed together, using the CLEAN-based deconvolution technique. A full explanation of this method can be found in \cite{gueth96} and \cite{gueth-guilloteau00}. As shown in Table~\ref{t:gl}, this process was limited by the rms of the less sensitive data set, and slightly degraded the angular resolution of the IRAM PdBI data but allowed the recovery of a large fraction of the extended emission.

Figure~\ref{f:compa} shows the map obtained from IRAM PdBI alone (Fig.~\ref{f:compa}a), from IRAM 30\,m alone (Fig.~\ref{f:compa}b), and the resulting map of the merging of the two data sets (Fig.~\ref{f:compa}c). On the latter, SiO emission appears bright at the location of the MDCs presented by \cite{louvet14} but also in regions where no MDCs were found. This is particularly striking on the northeastern section of the map, following the underlying filamentary structure.

\begin{table*}[htbp!]
\caption{Gain and loss through the merging processes}
\label{t:gl}
\centering
\begin{tabular}{cccc}
\hline
\hline
SiO(2--1) \& HCN(1--0)   & angular resolution (\asec)    & spectral resolution (km.s$^{-1}$)   & flux percentage (\%)  \\
\hline                                                                                                                                                
IRAM 30m                 & 29.9$\times$29.9              & 0.7                                 & 100                   \\
IRAM PdBI                & 3.14$\times$4.96              & 0.26                                & $\sim$10            \\
IRAM 30m+IRAM PdBI       & 3.28$\times$5.14              & 0.7                                 & $\sim$95            \\
\hline
\end{tabular}
\end{table*}

\section{Analysis and discussion}
\label{s:analysis} 

In this section, we analyze the SiO(2--1) and HCN(1--0) lines in detail to carry out a complete census of protostellar outflows detectable through line wings (see Sect.~\ref{s:outflows}). Then we quantify the contribution of each of the two components participating in the SiO emission profile (see Sect.~\ref{s:lowvelshock}).

\subsection{Protostellar outflows census}
\label{s:outflows}
 
\begin{figure*}[htbp!]
 \begin{center}
 \includegraphics[scale=0.06]{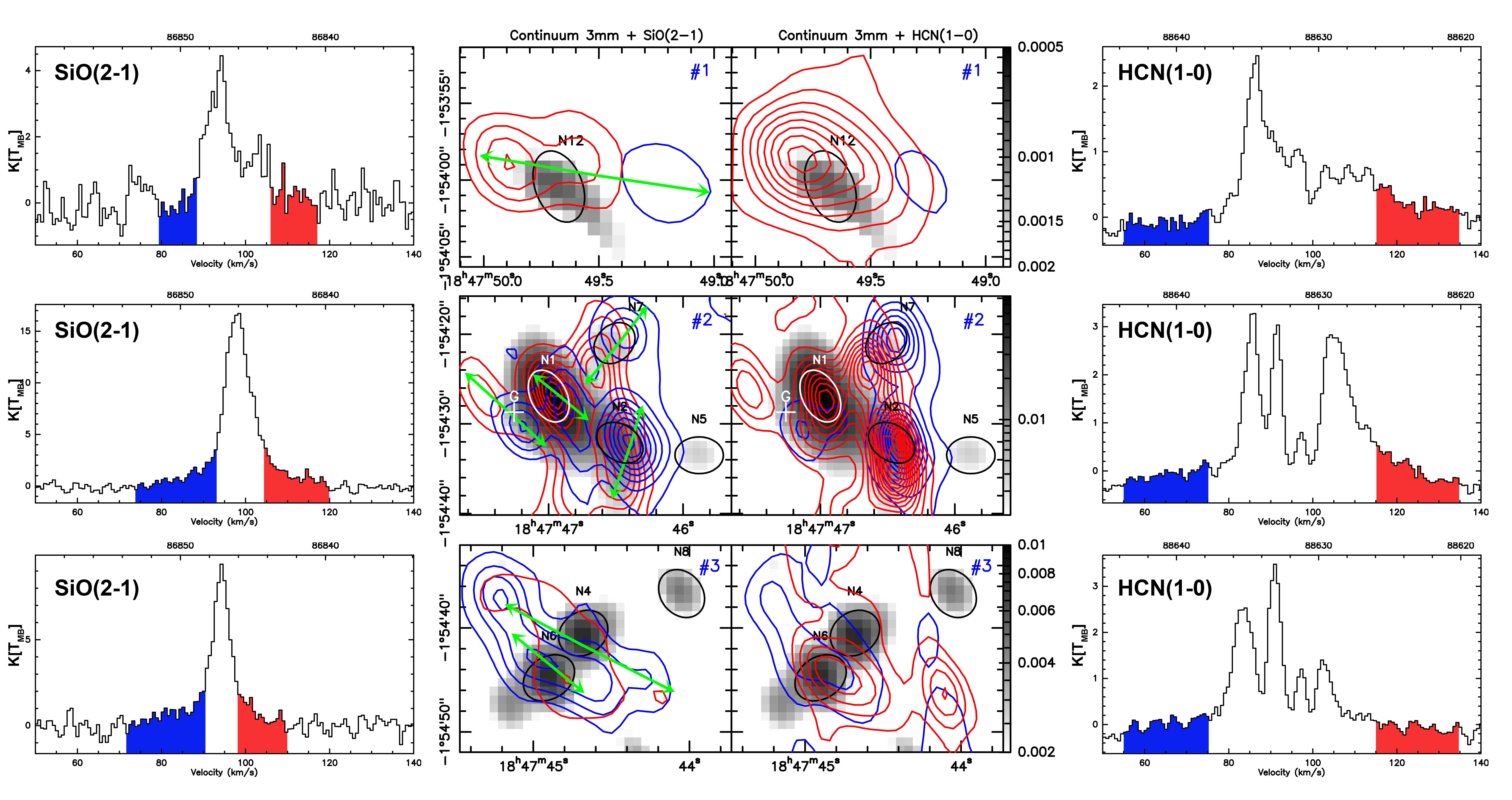}
  \caption{Seven outflows found in the \object{W43-MM1} ridge through SiO(2--1) and HCN(1--0) line wing integration. The top, middle, and bottom rows correspond to \#1, \#2, and \#3 boxes shown in Fig.~\ref{f:compa}c. The left panels show the mean SiO(2--1) emission line profiles in the areas corresponding to box \#1, \#2, and \#3, respectively. The right panels show the mean HCN(1--0) emission line profiles in the areas corresponding to box \#1, \#2, and \#3, respectively. In the SiO(2--1) and HCN(1--0) spectra, the channels filled with blue and red correspond to the velocity ranges used to define the blue wing and red wing, respectively. The middle panels show the 3\,mm continuum emission overlaid with SiO(2--1) contours on right panels and HCN(1--0) contours on left panels. The blue contours correspond to the blue line wings and the red contours correspond to red line wings. All contours start at 2$\sigma$ with 2$\sigma$ steps. The black and white ellipses highlight MDCs extracted from the interferometric continuum maps at 3\,mm presented in \cite{louvet14}, and the white cross on the middle panel locates source G of \cite{sridharan14}. Green arrows indicate outflow directions.}
 \label{f:flot}
 \end{center}
 \end{figure*}

In the stellar formation process, outflows develop as a result of angular momentum conservation during the collapse of a rotating core. Outflows are very strong indicators of early stage stellar formation. Protostellar ouflows are classically traced by CO lines and, to a lesser extent, through SiO, HCN, and HCO$^+$ lines \citep[e.g.,][]{bontemps96,gusdorf08a,duarte-cabral14}. Outflows are generally bipolar and have velocities that are offsetted with respect to those of protostars from which they are ejected. The relative velocities of shocked and/or entrained gas create line wings on each side of the rest frequency \citep{snell80}. We therefore integrate the blueshifted and redshifted wings of both the SiO(2--1) and HCN(1--0) lines to trace outflow lobes. The velocity of the filament in the local standard of rest is $\sim$98 km.s$^{-1}$ as defined with the N$_2$H$^+$(1--0) and SiO(2--1) lines by \cite{quang13}. Since HCN(1--0) is a four hyperfine component transition, whose first and last components are spaced by $\sim$20\,km.s$^{-1}$, we define spectral windows as 55--75\,km.s$^{-1}$ for the blue wing and 115--135\,km.s$^{-1}$ for the red wing. As the present study is focused on SiO emission, the spectral windows to define its blue- and redshifted wings are defined locally (see Fig.~\ref{f:flot}, left column). Indeed, the rest velocity of the SiO line varies by a few km.s$^{-1}$ along the filament. SiO and HCN lines, used together, enable us to complete a secure census of outflows in \object{W43-MM1}.

The search of outflows through blue and red components of both SiO(2--1) and HCN(1--0) lines was performed over the complete IRAM PdBI mosaic (see Fig.~\ref{f:compa}c). We found seven clear signatures of SiO(2--1) and HCN(1--0) outflows;  SiO often presenting a clearer view of outflow lobes (see Fig.~\ref{f:flot}). Six out of the seven outflows are driven by MDCs identified at 3\,mm by \cite{louvet14}: N1, N2, N4, N6, N7, and N12 with masses ranging from 20\,M$_{\odot}$ to 2000\,M$_{\odot}$ within $\sim$0.07\,pc. The presence of outflows leaves no doubt about their protostellar nature.
The last outflow appears to be associated with one of the seven cores extracted by \cite{sridharan14}  (see source G in see Fig.~ \ref{f:compa}c and Fig.~ \ref{f:flot}-\#2), of which four correspond to MDCs of \cite{louvet14}, while three are below their detection limit. Source \emph{G} has no published mass. From the Fig.~1 of \cite{sridharan14}, we estimate it to be $\sim$50$\pm$25\,M$_{\odot}$, when using a dust temperature and emissivity that is consistent with those taken in \cite{louvet14}. \cite{sridharan14} detected the outflow driven by N2 in their CO(3--2) image, but failed to disentangle those of N1 and N7 MDCs.

We use the SiO and HCN peaks of the blue- and redshifted outflow lobes to define outflow directions. Blue- and redshifted lobes are not perfectly aligned with MDC locations (see Fig.~\ref{f:flot}), possibly because of MDCs subfragmentation, uneven morphology of the surrounding environment, and/or chemistry effects in the shock. The average projected separation between outflow lobes is $\sim$0.20\,pc and, surprisingly, the outflow arising from the most massive MDC N1, is the smallest ($<$\,0.08\,pc) at the limit of our resolution.
Either the outflow arising from W43-N1 MDC is mostly along the line of sight or it is too young and/or too embedded to have traveled far.
The outflows, driven by the MDCs N1, N4, N6 of \cite{louvet14} and that from the source G of \cite{sridharan14}, are aligned with the SiO(2--1) emission shown in Fig.~\ref{f:compa}c (see Fig.~\ref{f:flot}). They are however relatively compact and cannot explain the $\sim$5\,pc-long SiO emission by themselves. In the northeastern part of the ridge, no known protostellar activity can account for the SiO(2--1) emission either. Indeed, \cite{louvet14} did not detect any dense core above their $\sim$1.5\,M$_\odot$, $3\sigma$ sensitivity limit, and the present study does not identify any evidence of a high-velocity outflow in this area.

\subsection{The two components of the SiO emission line}
\label{s:lowvelshock}

\begin{figure}[htbp!]
\centerline{
\includegraphics[scale=0.35]{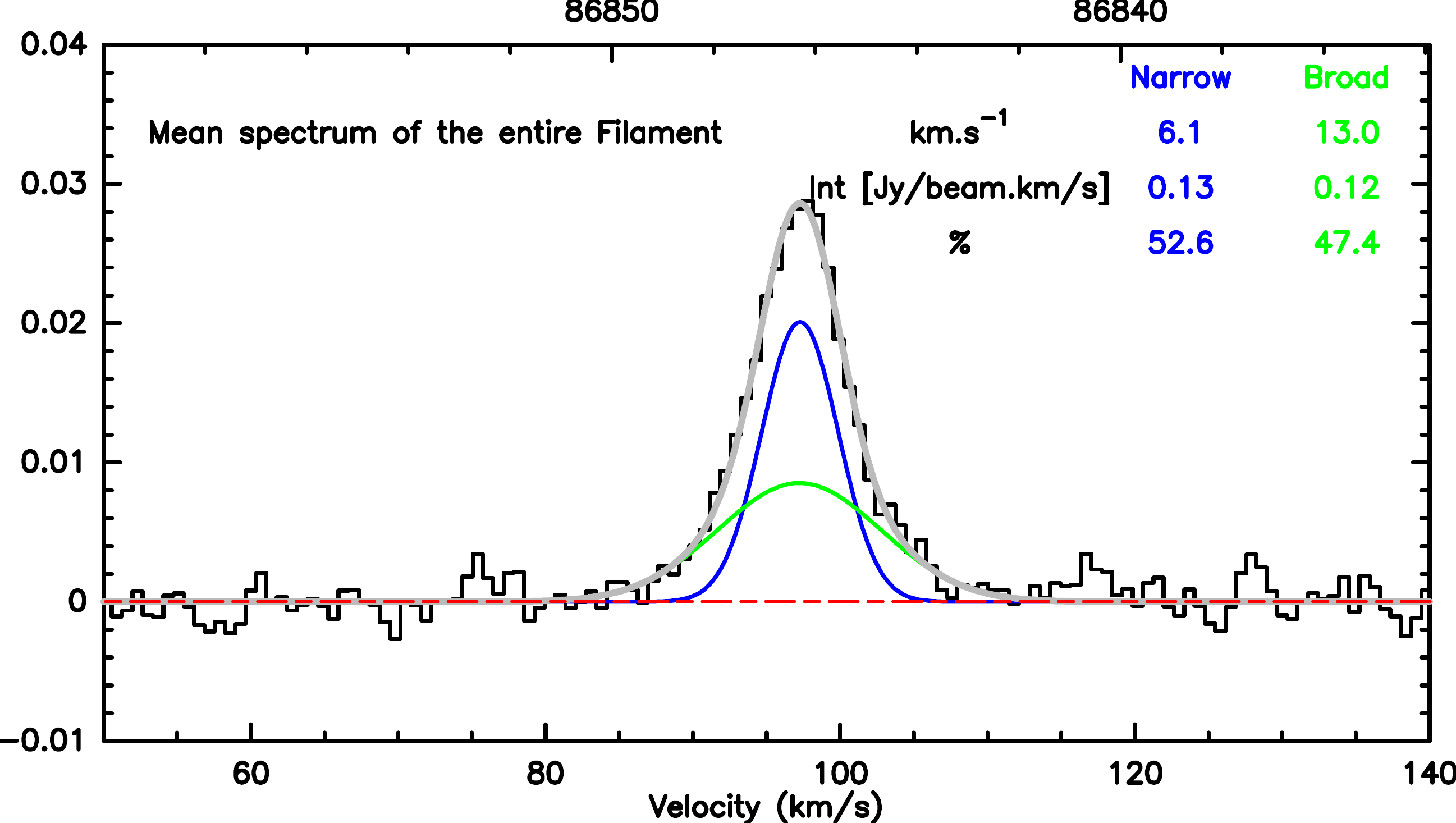}}
\caption{Combined IRAM PdBI and 30 m SiO(2--1) spectrum (black histogram) averaged over the entire area shown in Fig.~\ref{f:compa}c. It is fitted by the sum of a narrow (blue curve) and a broad (green curve) Gaussian components are shown with the gray curve. The red dashed line indicates the zero level baseline.}
\label{f:filament}
\end{figure}

Figure~\ref{f:filament} presents the SiO(2--1) line spectra averaged over the entire area presented in Fig~\ref{f:compa}c. It suggests, as for studies conducted in other high-mass star-forming regions \citep{jimenez10,sanhueza13,duarte-cabral14}, that the SiO line profile consists of the sum of two Gaussian components. The fit of the SiO line profile (see Fig.~\ref{f:filament}) is composed of a broad component ($\sim$13\,km.s$^{-1}$) providing $\sim$45\% of the total SiO emission and a narrow component (FWHM\,$\sim$6\,km.s$^{-1}$) providing $\sim$55\% of the emission. 

\begin{figure*}[htbp!]
\centerline{
\includegraphics[scale=0.7]{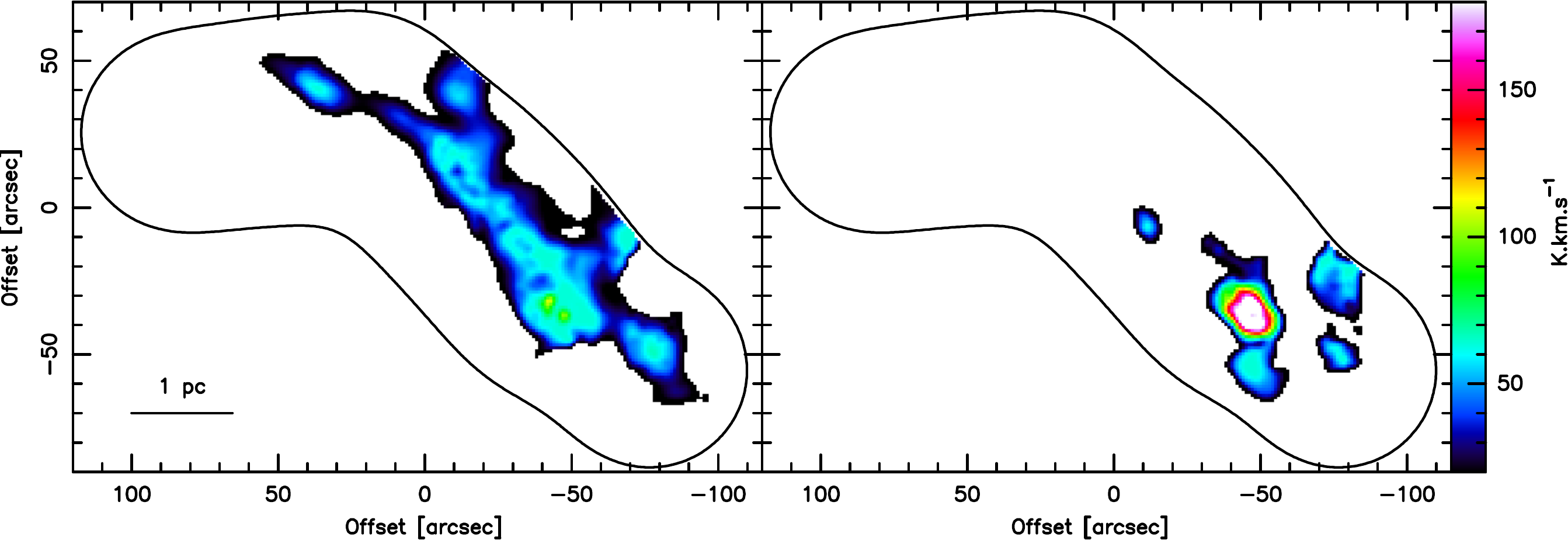}}
\caption{Spatial distribution of the narrow ({\bf left panel}, $<$\,12 km.s$^{-1}$) and wide ({\bf right panel}, $>$\,12\,km.s$^{-1}$) components fitted to the SiO lines of the \object{W43-MM1} ridge, shown from the integrated fluxes of Gaussian fits of these two components. The back contour highlights the area covered by the IRAM PdBI observations.}
\label{f:location}
\end{figure*}

Figure~ \ref{f:location} presents the spatial distribution of the narrow and broad components. This decomposition results from a double-component Gaussian fitting of the spectra measured toward each pixel. The three parameters (FWHM, v$_{\rm lsr}$, and peak flux) of each Gaussian were left free. A double-component fit was considered successful if each component had a peak flux $>\,3\sigma$ and a FWHM size\,$<\,60$\,km.s$^{-1}$. Moreover, we imposed the difference between the v$_{\rm lsr}$ of the two Gaussians to be smaller than the FWHM of the broader component, so that the Gaussians would overlap. If those conditions were not fulfilled, we would perform a single Gaussian fit component to the spectra. The single Gaussian fit was kept if its peak flux was above 3$\sigma$. The area covered in Fig.~ \ref{f:location} is smaller than Fig.~\ref{f:compa}c because of these criteria. The left panel shows the distribution of the emission associated with the narrow component, with FWHM $<12$\,km.s$^{-1}$, which follows the underlying filamentary structure over $\sim$5\,pc long. The right panel, which is associated with the broad component with FWHM $>12$\,km.s$^{-1}$, pinpoints all the positions where outflows have been detected. 

To complement the spatial separation made in Fig.~\ref{f:location}, we hereafter characterize the two components of the emission, quantify their local, relative contributions to the total SiO(2--1) emission, and debate their different origin. To do so, we focus on eight areas along the filament (labeled from 1 to 8), which are defined in Fig.~\ref{f:location}c.

\subsubsection{Broad component only}
\label{ss:broad}

Of the eight positions selected, only position 4 may be fitted by a single broad Gaussian (see Figs.~\ref{f:location} and~\ref{f:raies}). This position corresponds with the location of the N12 MDC, where an outflow has been extracted (see Sect.~\ref{s:outflows}, and Fig.~\ref{f:flot}). It is well fitted by a Gaussian with a FWHM of $\sim$11\,km.s$^{-1}$.

\subsubsection{Narrow component only}
\label{ss:narrow}

\begin{figure*}[htbp!]
\centerline{
\includegraphics[trim=0cm 0cm 0cm 0cm, scale=0.25]{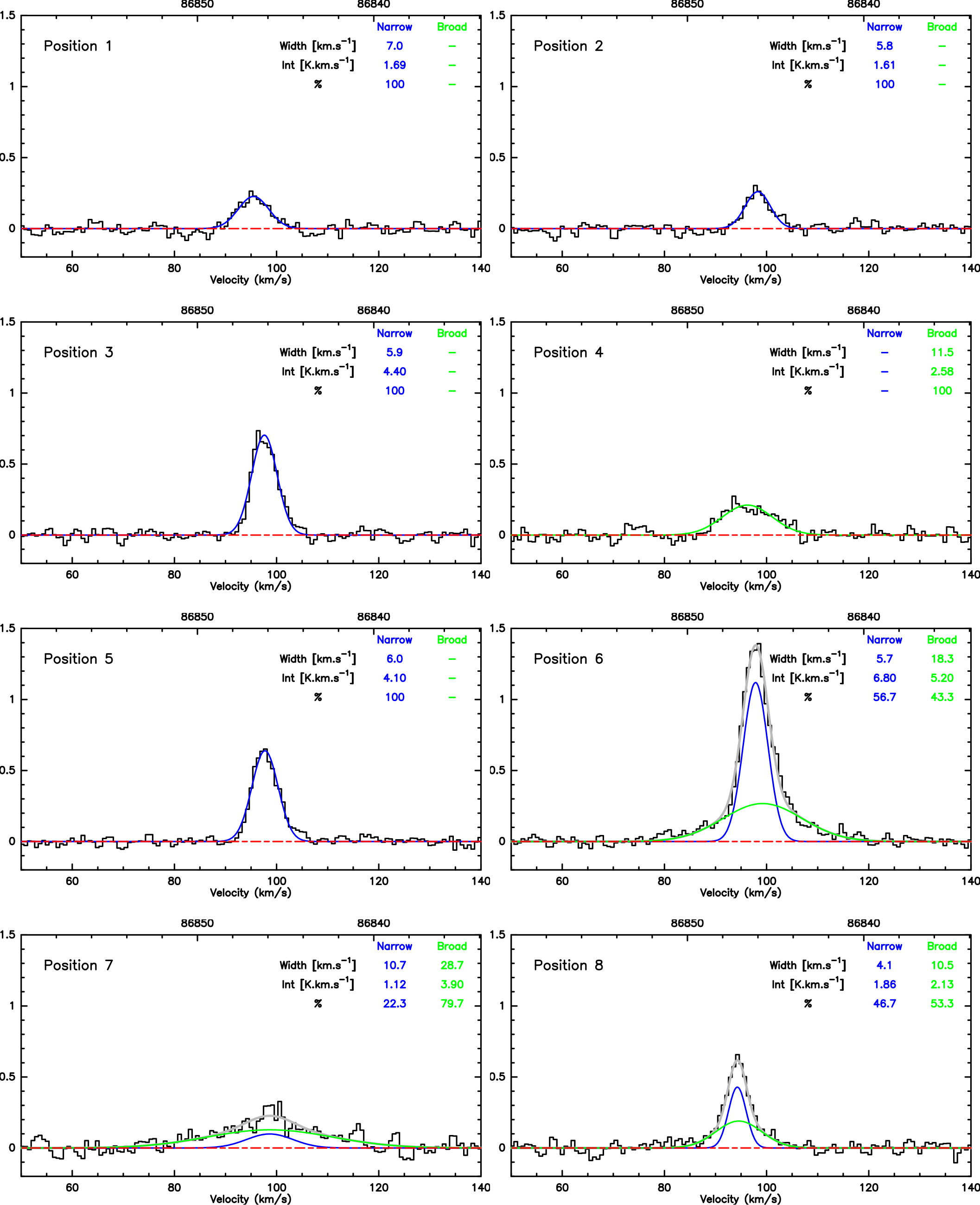}}
\caption{SiO(2--1) molecular lines, averaged for the eight areas shown in Fig.~\ref{f:compa}c with white circles. These lines are fitted by the sum of narrow (blue curve) and broad (green curve) Gaussian components shown by the gray curve. The red dashed line shows the zero level baseline.}
\label{f:raies}
\end{figure*}

Positions 1, 2, 3, and 5 (see Figs.~\ref{f:location}-\ref{f:raies}), which correspond to the northeastern extended emission of SiO ($\sim$3\,pc long), display a very homogeneous emission. If the peak emission slightly increases toward the mosaic center (from $\sim$3\,K at position 1 to $\sim$7\,K at position 5), the width of the narrow Gaussian profile is extremely regular, $\sim$6\,km.s$^{-1}$. 

\subsubsection{Double component}
\label{ss:both}

Position 6 and 8 require both one broad and one narrow component to be properly fitted. Toward position 8, the broad component is $\sim$10.5\,km.s$^{-1}$, which is very similar to that of position 4. Since outflows associated with MDCs, labeled N4 and N6, are clearly detected at position 8, the broad Gaussian component is most probably associated with star formation in that area. The narrow Gaussian component, however, is more interesting. Accounting for almost half of the total SiO emission, it only has a FWHM of $\sim$4\,km.s$^{-1}$, which is consistent but narrower than the narrow Gaussian component in the northeastern part of the ridge.

As for position 6, the broad ($\sim$20 km.s$^{-1}$) component is associated with star formation linked to the outflows of the MDCs labeled N1, N2, N7, and G. The linewidth is broader than at position 4 ($\sim$20\,km.s$^{-1}$ instead of $\sim$11\,km.s$^{-1}$), which is most likely due to the multiple outflows (four) at position 6 (Sect.~\ref{s:outflows}). The narrow component in this central region, with a FWHM of $\sim$6\,km.s$^{-1}$, has the same characteristics as the narrow component located to the northeast and southwest of position 6.

\smallskip 

Position 7 also needs a double Gaussian component fit. Nevertheless, its broad and narrow components show $\sim$60\,\% broader profile than star-forming regions of the main filament (positions 6 and 8). A cluster of low-mass protostellar outflows could explain the broader line component, while multiple low-velocity shocks could explain the narrow line component. Further observations are necessary to determine the origin of the emission at this position.

To summarize, the SiO line is fitted by a broad component at locations where star formation is known to occur and  a narrow component with a $\sim$6\,km.s$^{-1}$ FWHM everywhere.

\subsubsection{Origin of the narrow component}
\label{ss:ori}

Most of the extended ($\sim$5\,pc long from position 1 to 8) narrow component of the SiO line emission likely has the same origin throughout the map. The narrow component contribution, participating from 45\% and up to 100\% to the SiO(2--1) emission, shows a regular line width of $\sim$6\,km.s$^{-1}$ along the 5\,pc projected distance where the SiO is detected. Since protostellar activity cannot account for the SiO emission at positions
1, 2, 3, and 5 (see Sect.~\ref{s:outflows}), we conclude that most of the narrow profile can only be associated with low-velocity shocks of the ISM, not from outflows.

The physical process generating those low-velocity shocks is not settled. More observations are necessary to constrain the process from a few possibilities. The shocks may originate from the formation of the ridge by large-scale cloud-cloud collision. To prove that origin, one would need to trace the gas from those clouds at large scale and confirm that they collide at low velocity. However, the mass accumulated in the ridge exceeds 10$^4$\,M$_{\odot}$. 
The ridge thus gravitationally dominates its environment. As proposed by \citet{vazquezsemadeni05}, \citet{hartmann12}, and \citet{smith13} in numerical simulations, the gravitational potential could generate colliding flows funneling mass from large scales down to the scale of ridges.
To confirm that origin, one would need to trace those colliding flows at large scale and compare their terminal velocities with respect to that of the ridge. In the latter case, shocks would occur at the ridge edges. We discuss the possibility of forming the amount of SiO observed from low-velocity shocks and the detailed physicochemical process that allows these shocks to release SiO into the gas phase in Sect.~\ref{s:model}.


\section{Low-velocity shocks in numerical models}
\label{s:model}

The detection of extended emission of SiO associated with relatively narrow linewidths that are unrelated to protostellar jets or outflows is the most puzzling aspect of our study. This puzzle is best illustrated in positions 1, 2, 3, and 5, where star formation activity is not present. In this section, we present a grid of low-velocity ($v_{\rm s} \leq 20$~km~s$^{-1}$) shock models with the aim to account for this SiO emission. We base our discussion on the use of the SiO(2--1) emission line only in positions 3, 5, and 8. The other positions will be explored in a forthcoming article that will thoroughly describe the emission of SiO lines in low-velocity shock models.

\subsection{The grid of shock models}
\label{sub:ogosm}

We calculate the SiO(2--1) integrated intensity over a grid of models computed with the Paris-Durham 1D shock model. The global architecture of this model was presented in \citet{flower15}, but we use a slightly different version that was introduced in \citet{lesaffre13}. This version allows us to run mildly irradiated shock models. These are shocks subject to an ultraviolet radiation field provided that the irradiation does not significantly exceed the interstellar radiation field (ISRF). The measure of this radiation field is a so-called $G_{\rm 0}$ factor, which is a scaling factor with respect to the ISRF defined by \citet{draine78}. In order to calculate SiO(2--1) integrated intensities from our models, we then use a post-processing treatment based on the large velocity gradient approximation, following a method already used in \citet{gusdorf08a, gusdorf08b} and more recently \citet{gusdorf15}.

We cover the following input parameters: preshock density of 10$^3$, 10$^4$, 10$^5$, 10$^6$\,cm$^{-3}$ ; shock velocity $v_{\rm s}$ from 4 to 20\,km.s$^{-1}$ in steps of 2\,km.s$^{-1}$, magnetic field parameter $b = 1$ (defined by $B$($\mu$ G) = $b \times [n_{\rm H}$ (cm$^{-3}$)]$^{1/2}$, where $B$ is the intensity of the magnetic field transverse to the shock direction of propagation), and the external radiation field $G_{\rm 0} = 0$, 1. Our grid only contains stationary C-type models, given the adopted preshock density and magnetic-field parameter (\citealt{lebourlot02,flower-pdf03}). Our grid does not include the effects of grain--grain interactions, as presented in \citet{guillet11} and \citet{anderl13}, as these effects  should only be taken into account when a high preshock density ($n_{\rm H} > 10^4$~cm$^{-3}$) is combined with a high shock velocity ($v_{\rm s} > 20$~km~s$^{-1}$). The input parameters that we investigate are summarized in Table~\ref{table-ag1}. Our choices were based on the following observational constraints:

\paragraph{The preshock density} value that is used in our models to account for the observations in position 1 was estimated in \cite{louvet14}. The column density was derived from \emph{Herschel} continuum data first presented in \citet{quang13}. Further  geometrical assumptions were made to compute the cloud volume density, after having subtracted the foreground and background emissions. The corresponding cloud volume density is 10$^4$~cm$^{-3}$ (for position 3) and $\sim$\,5$\times10^4$~cm$^{-3}$ (for positions 5 and 8). Therefore, we only present the part of our models grid that covers a preshock density of 10$^4$~cm$^{-3}$. In fact, at considered shock velocities, this preshock density value converts into a postshock density value of a few (less than 10) 10$^4$~cm$^{-3}$, corresponding to the observational constraint inferred from \emph{Herschel}. A more complete grid will be presented in a forthcoming publication, which will include both more preshock density values (10$^3$, 10$^5$, and 10$^6$~cm$^{-3}$) and present predictions for SiO lines up to $J_{\rm up} = 20$. 

\paragraph{Shock velocity.} The SiO emission is generated in the shocked gas. The width of the SiO(2--1) emission line therefore constitutes an indication of the shock velocity. Given the number of observational constraints, namely the SiO(2--1) integrated intensity, we restrain our goal to fitting this quantity by means of one single 1D shock structure. The width of the SiO(2--1) line profile derived in Sect.~\ref{ss:narrow}, of FWHM ranging from $\sim$\,4 km.s$^{-1}$ to $\sim$\,7 km.s$^{-1}$, is consistent with that proposed earlier by \cite{quang13}. This velocity translates to a shock velocity of $\sim$\,8-14 km.s$^{-1}$ if a single shock is observed along the line of sight. In the context of one-dimensional modeling, adopted here, the shock is assumed to be seen face-on. However, it is possible that shocks propagate in an already moving medium or that the flow is orientated toward the plane of the sky, resulting in lower or higher shock velocities, respectively. For this reason, we designed a grid covering shock velocities from 4 to 20~km~s$^{-1}$. 

\paragraph{Magnetic field parameter.} The magnetic field is measured along the line of sight of the MDC N1 (see Fig.~\ref{f:compa}-c and Fig.~\ref{f:flot}) to be $\sim$6\,mG \citep{sridharan14}. The volume density of the MDC N1, previously derived as $\sim$10$^8$\,cm$^{-3}$ \citep{louvet14}, would roughly correspond to a $b$ value equal to one for this shock component. We keep this value ($b$=1) over our grid of calculations, which also corresponds to a standard value for molecular clouds of the interstellar medium (e.g., the review by \citealt{crutcher12}). 

\paragraph{Radiation field.} At the high cloud volume densities reached within the filament (up to a few $10^6$\,cm$^{-3}$) and despite the presence of a starburst cluster located a projected distance of 5--10\,pc from the \object{W43-MM1} filament \citep{blum99,motte03}, the effective radiation level within the filament is weak. The radiation of the WR and O-type cluster, estimated to be as high as $\sim10^4$~$G_{\rm 0}$ (N. Schneider, private communication), quickly decreases when penetrating the high density of the \object{W43-MM1} filament. More specifically, the intensity of the radiation field is negligible ($<$0.1~$G_{\rm 0}$) when the column density reaches a few $10^{21}$~cm$^{-2}$. Nevertheless, for completeness, we chose to present shock models without any irradiation ($G_{\rm 0}$ = 0), or subject to the standard ISRF proposed by \citet{draine78} ($G_{\rm 0}$ = 1). The models that we ran are subject to a UV radiation coming from the side; the $N_{\rm H_2}$ and $A_{\rm v}$ values are constant.

\begin{table}[htbp!]
\caption{Explored parameters in our grid of shock models. The stationary run determines the (preshock) initial abundances of our models.}
\addtolength{\tabcolsep}{-5pt}
\label{table-ag1}
\begin{center}
\begin{tabular}{| l | c |}
\hline
Parameter                               & Explored values                                                                 \\
\hline
\hline                                                                                                                            
$n_{\rm H}$ [cm$^{-3}$]                 & 10$^3$\,$^{\rm a}$ ; 10$^4$ ; 10$^5$\,$^{\rm a}$ ; 10$^6$\,$^{\rm a}$           \\
$v_{\rm s}$ [km.s$^{-1}$]               & 4 ; 6 ; 8 ; 10 ; 12 ; 14 ; 16 ; 18 ; 20                                         \\
$G_{\rm 0}$ (no unit)$^{\rm b}$         & 0 ; 1                                                                           \\
\hline
\hline
\small{Free stationary-run Si (\%)}     &  1, 10                                                                          \\
\small{Form of free stationary-run Si}  &  gas-phase Si$^{\rm c}$, grain-mantles SiO$^{\rm d}$                            \\
\hline
\end{tabular}
\end{center}
$^{\rm a}$: to be presented in a forthcoming publication.                                                                 \\
$^{\rm b}$: multiplicative factor with respect to \citet{draine78}'s ISRF.                                                \\
$^{\rm c}$: so-called SiG scenario, see Sect.~\ref{sub:naosioc}.                                                                                       \\
$^{\rm d}$: so-called SiM scenario, see Sect.~\ref{sub:naosioc}.
\end{table}

\subsection{Silicon chemistry}
\label{sub:naosioc}

The most important aspect in this grid is related to the assumptions we adopted on the silicon chemistry. The following stages are classically considered when modeling the SiO emission. In the preshock phase, the silicon-bearing material is locked up in the cores of the interstellar grains in the form of silicates, such as olivine, forsterite, or fayalite (e.g., \citealt{may00}). When the shock wave propagates, the grains are charged (e.g., \citealt{flower-pdf03}). They consequently undergo collisions from the neutral species in the gas phase, which leads to sputtering of the grain cores and release of silicon in the gas phase (e.g., \citealt{schilke97}). The gas-phase silicon then reacts either with O$_2$ (e.g., \citealt{lepicard01}) or with OH (e.g., \citealt{riverosantamaria14}) to produce SiO. Next, SiO is removed from the gas phase in the postshock region through gas-phase chemistry (and conversion in SiO$_2$) or through adsorption on the grain surface (\citealt{gusdorf08a}). Mostly because of the sputtering yields, it has been shown that this formation path for SiO could be efficient. It would generate levels of emission that are compatible with the observations only in shocks with a velocity greater than 25~km~s$^{-1}$. 

But the remaining fraction of the silicon abundance, up to 10\% of the solar type value, is free and can belong either to the gas phase or to grain mantles (e.g., \citealt{jenkins09}). This view is consistent with results obtained in shocks associated with the formation of low-mass (in L1157, \citealt{gusdorf08a}, or BHR71, \citealt{gusdorf15}) or high-mass stars \citep{leurini14}. The quantity and form of the free silicon is not known, as it depends on the physical conditions of the preshock gas (temperature, density, and radiation field). We therefore chose to place 1\% or 10\% of free silicon in the form of neutral silicon in the gas phase (SiG scenario), or SiO in the mantles of the interstellar grains (SiM scenario) in our preshock conditions (which are in stationary state). The remaining fraction of the solar value (99\% and 90\%, respectively) is maintained in the core of interstellar grains. These assumptions are given in Table~\ref{table-ag1}.

The number of models in the fraction of our grid presented here is consequently [2 (free Si fraction assumption)] $\times$ [2 (free Si form assumptions)] $\times$ [1 (preshock density value)] $\times$ [2 (irradiation strength)] $\times$ [9 (velocity values)], or 72 models. The current work presents two improvements over the previous silicon chemistry studies in low-velocity shocks presented in \citet{quang13} as follows:
\begin{itemize}
\item[$\bullet$] Our grid is more extended in terms preshock silicon fraction; it is also tighter in shock velocity, and more precise in terms of $b$ factor, as this one is estimated based on smaller scale measurements. It is slightly different in terms of $G_{\rm 0}$ values, as the relatively high densities derived in \citet{louvet14} indicate that the effective radiation within the shock structures correspond to lower $G_{\rm 0}$ factor.
\item[$\bullet$] Our work presents direct model -- observation comparisons, in the sense that integrated intensities are calculated in our models, and that filling factor corrections are applied to the observations. This way, no assumption related to the SiO column density estimation intervenes between the models and observations. 
\end{itemize}

\subsection{Grid fragment results}
\label{sub:gfr}

\begin{figure}[h]
\centering
\includegraphics[width=0.45\textwidth]{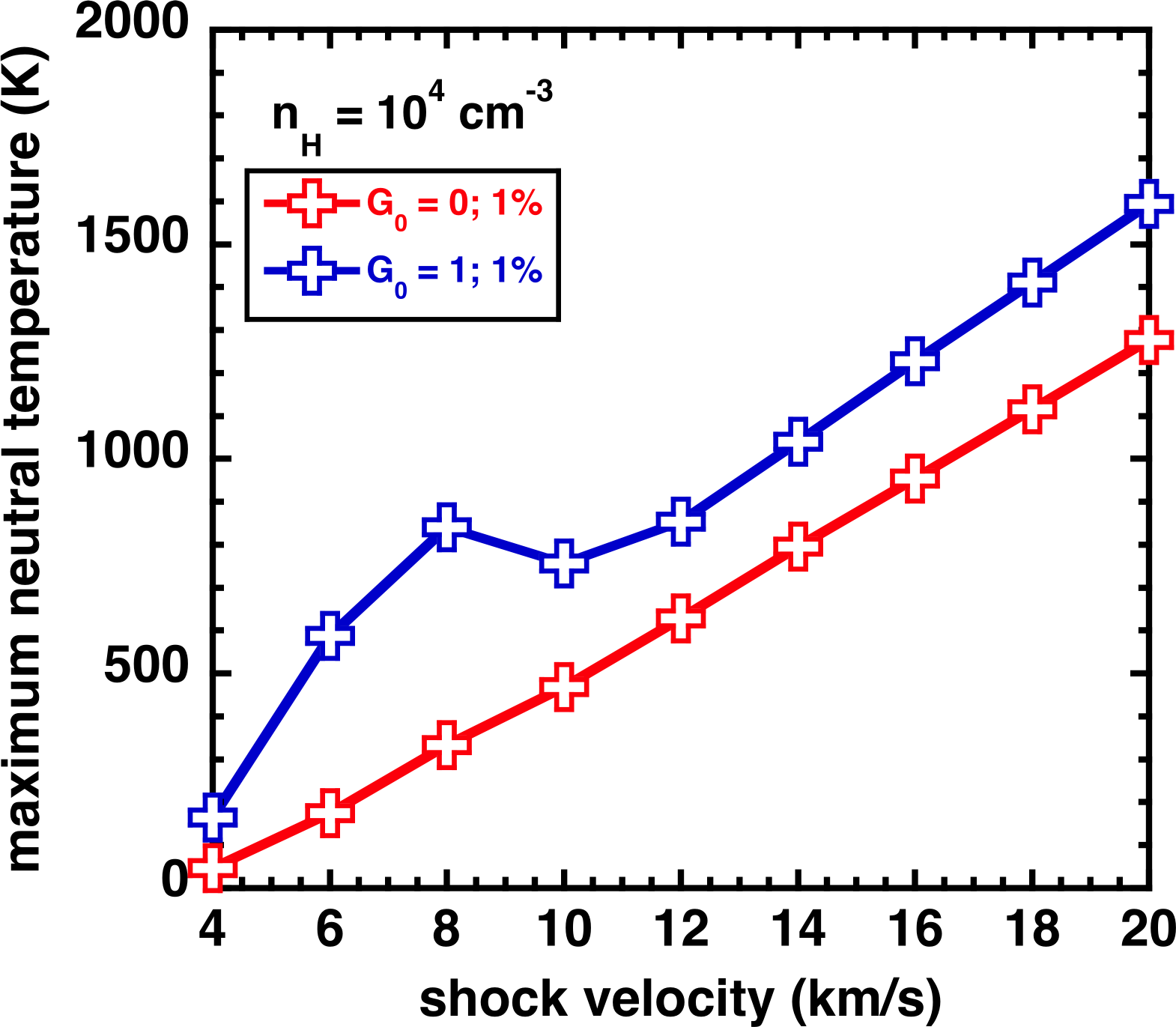}
\caption{Maximum neutral temperature reached in our models against the shock velocity. This value depends very weakly on the fraction of preshock free silicon and on the related chemistry. It does, on the other hand, depend on the $G_{\rm 0}$ value.}
\label{figure2ant}
\end{figure}

\begin{figure*}
\centering
\includegraphics[width=0.8\textwidth]{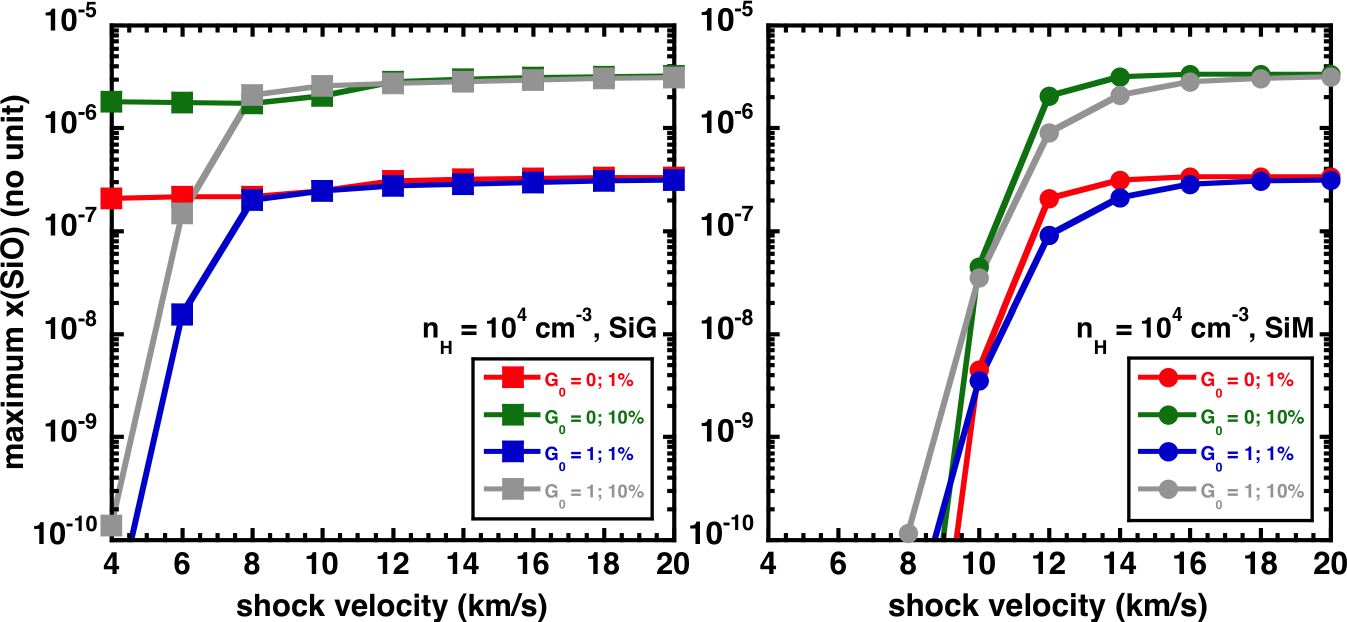}
\caption{Both panels show the maximum local fraction of SiO reached in each of the shock models, shown against the shock velocity. The maximum local fraction of SiO is defined as x(SiO)=$n_{\rm SiO}/n_{\rm H}$. The left and right panels correspond to the SiG (square symbols) and SiM (circle symbols) scenarios (see text), both for a preshock density $n_{\rm H}=10^4$\,cm$^{-3}$. The color code is indicated in the panels, where the $G_{\rm 0}$ (= 0 or 1) value and fraction of preshock free silicon (= 1 or 10 \%) is shown for each model.}
\label{figure3ant}
\end{figure*}

In order to understand the results of our grid, Figs.~\ref{figure2ant} and~\ref{figure3ant} present the maximum neutral temperature and maximum local abundance, respectively, of SiO reached in our models. 

The maximum neutral temperature value insignificantly depends on silicon chemistry or abundance. It does, on the other hand, depend on the external irradiation. This is why we chose to show it only for $G_{\rm 0} = 0, 1$, with 10\% of free preshock silicon in the gas phase. We find that the maximum temperature linearly increases with shock velocity when $G_{\rm 0} = 0$. The evolution of the maximum neutral temperature when $G_{\rm 0} = 1$ is more complex. The radiation heats the medium directly through photoelectric effect, or indirectly through the dissociation of molecules that normally act as cooling agents (such as H$_2$). This explains globally why the blue curve is above the red curve in Fig.~\ref{figure2ant}. However the variation of the blue curve is not monotonic. For C-type shock models, ions interact with the magnetic field. The heating of the gas mostly comes from the friction between ions and neutrals, and the charged fluid is dominated by C$^+$. C$^+$ direct recombination with electrons is inefficient and C$^+$ indirectly recombines through the chemical chain of formation of CH$_{\rm n}^+$ with efficient dissociative recombination for each CH$_{\rm n}^+$. The first reaction of this chain, C$^+$ + H$_2 \rightarrow $CH$^+$, is subject to a strong energy barrier. However, since it involves an ion and a neutral species, the drift velocity between ions and neutrals (proportional to the shock velocity) can overcome this barrier. When the shock velocity is above 8\,km.s$^{-1}$, the reaction chain is activated and C$^+$ recombines, leaving the less abundant cation S$^+$ as the main charge carrier. The ion-neutral friction heating hence diminishes and the maximum neutral temperature in the shock steps down to a value closer the nonirradiated case, where S$^+$ is also the main charge carrier. This explains the different slopes of the blue curve below and above 8\,km.s$^{-1}$. We note that this process does not affect the maximum temperature in nonirradiated shock models, for which S$^+$ and not C$^+$ is the main carrier of the charge.

{We must then interpret Fig.~\ref{figure3ant}. We distinguish three cases:
\begin{itemize}
\item Free preshock silicon in the gas phase (SiG), $G_{\rm 0} = 0$ scenario. In this case, our stationary models predict that almost all the silicon is in the form of SiO in the preshock phase. This fraction does not depend on the shock velocity. During the shock propagation, the SiO abundance only decreases in the postshock because of adsorption onto grains when the temperature drops. The SiO maximum local abundance is subsequently the preshock value and does not significantly vary with the shock velocity. As expected, it does depend on the fraction of free silicon; the SiO maximum abundance is ten times higher when 10\% of free silicon rather than 1\% is placed in the preshock gas.
\item SiG, $G_{\rm 0} = 1$ scenario. In this scenario, our stationary models predict that Si$^+$ is the dominant form of silicon in the preshock phase (over Si and SiO). After the temperature rise, a fraction of Si$^+$ is converted to SiO. This fraction, the local abundance of SiO is regulated by its destruction when it interacts with C$^{+}$, whose large abundance is caused by the larger $G_{\rm 0}$ factor. When the shock velocity increases, C$^+$ gets consumed in the high-drift chemistry (it is converted to CH$_{\rm n}^+$, see the above discussion on temperature), so less SiO is destroyed. This explains the increase in maximum SiO abundance with shock velocity seen in Fig.~\ref{figure3ant} until the shock velocity reaches 8\,km.s$^{-1}$, where it reaches a plateau. We note that above this threshold of 8\,km.s$^{-1}$, the maximum abundance of SiO does not depend on the radiation field value, and that the maximum SiO abundance is ten times higher when 10\% of free silicon is placed in the preshock rather than 1\%.
\item Free preshock silicon in grain mantles (SiM), $G_{\rm 0} = 0, 1$ scenario. The abundance of silicon-bearing material in the preshock gas phase is negligible. In fact, a great majority of the silicon belongs to the grains, whether it is to their core or mantle. The evolution of the maximum fraction of SiO abundance in the gas phase with the shock velocity reflects the existence of a velocity threshold associated with the sputtering processes that affect the mantles and lead to the direct release of SiO in the gas phase (e.g., calculations in \citealt{flower-pdf94} building on earlier results on sputtering from \citealt{barlow78}). The sputtering process is very efficient, so that all the SiO initially placed in the grain mantles is injected in the gas phase for $v_{\rm s} \geq 12$\,km.s$^{-1}$. It is only removed from the gas phase in the postshock by adsorption when the gas is cold enough when $G_{\rm 0} = 0$, and by reaction with C$^+$ when $G_{\rm 0} = 1$. The destruction process does not affect the maximum local abundance.
\end{itemize}

\begin{figure*}
\centering
\includegraphics[width=0.8\textwidth]{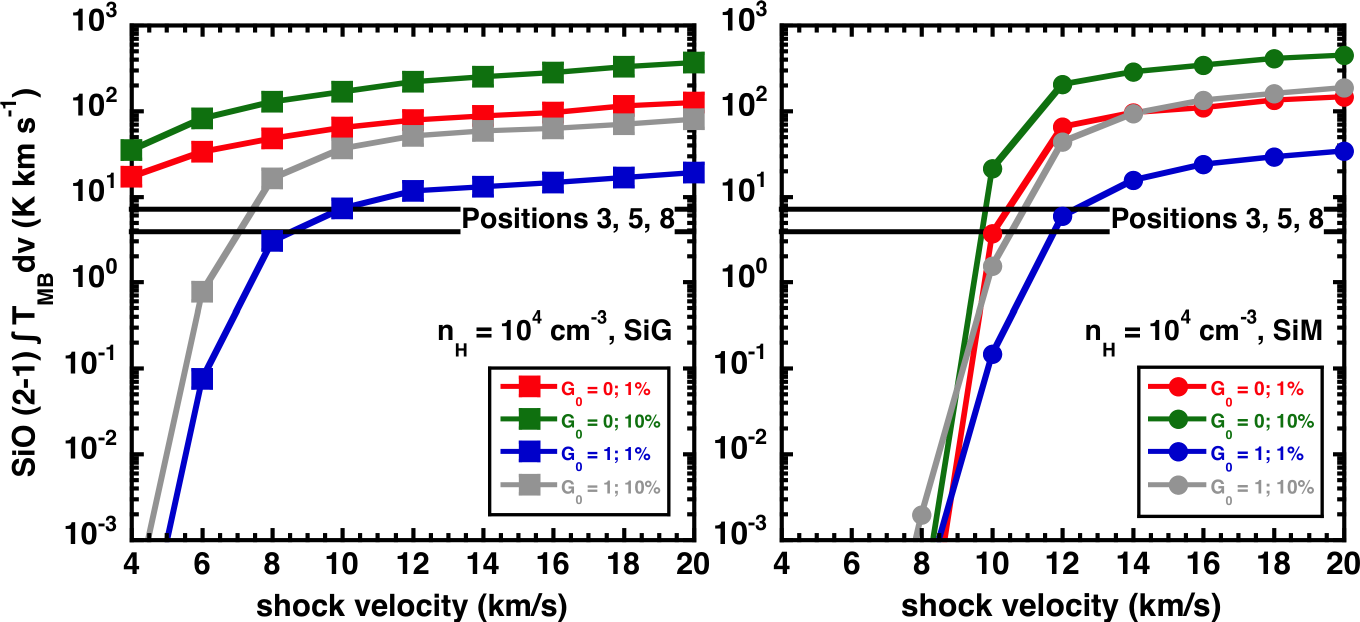}
\caption{Integrated intensity of the SiO(2--1) transition against the shock velocity calculated over the $n_{\rm H} = 10^4$~cm$^{-3}$ fragment of our grid of models (colored symbols), and compared to the observations (thick, horizontal black line, obtained in position 1, with thinner lines corresponding to $\pm 15$\% uncertainty; see Tables~7 and 8). The left and right panels correspond to the SiG (square symbols) and SiM (circle symbols) scenarios (see text). The color code is indicated in the panels, where the $G_{\rm 0}$ (= 0 or 1) value and fraction of preshock free silicon (= 1 or 10 \%) is shown for each model.}
\label{figure1ant}
\end{figure*}

\subsection{Comparison with observations}
\label{sub:cwo}

The results of our grid in terms of SiO(2--1) integrated intensity are shown in Fig.~\ref{figure1ant}. The $n_{\rm H} = 10^4$~cm$^{-3}$ fragment of our grid of models is shown. The left-hand column represents the SiG models, whereas the right-hand column represents the SiM models. The color code is indicated in the panels and one model point is shown for each shock velocity (from 4\,km.s$^{-1}$ to 20\,km.s$^{-1}$). Overall, contrary to the maximum local abundance of SiO, the SiO (2--1) integrated intensity is never proportional to the free, preshock silicon quantity. The results from our models can be interpreted as follows:
\begin{itemize}
\item SiG, $G_{\rm 0} = 0$ scenario. The SiO (2--1) integrated intensity slightly increases with shock velocity. This is a consequence of the maximum temperature increase seen in Fig.~\ref{figure2ant}, as the maximum local abundance of SiO does not depend on the shock velocity.
\item SiG, $G_{\rm 0} = 1$ scenario. The SiO (2--1) integrated intensity is lower than in the $G_{\rm 0} = 0$ case because the SiO emitting layer is thinner; the overall shock layer is smaller because of the higher ionization fraction, and the layer fraction where SiO exists is thinner because of its destruction by C$^+$. The increase with shock velocity is a consequence of the combination of the temperature and abundance increase as seen in Figs.~\ref{figure2ant}-\ref{figure3ant}.
\item SiM, $G_{\rm 0} = 0$ scenario: the increase of the SiO (2--1) integrated intensity with shock velocity is again a consequence of the combination of the temperature and abundance increase revealed in Figs.~\ref{figure2ant}-\ref{figure3ant}. Above the sputtering yield at 12\,km.s$^{-1}$, the level of emission is completely comparable to that generated in the \{SiG, $G_{\rm 0} = 0$\} scenario. This is because the SiO abundance profiles through the shock layers are similar for these two categories of models.
\item SiM, $G_{\rm 0} = 1$ scenario. The increase of the SiO (2--1) integrated intensity with shock velocity is a consequence of the combination of the temperature and abundance increase revealed in Figs.~\ref{figure2ant}-\ref{figure3ant}. However, the SiO (2--1) integrated intensity is lower than in the $G_{\rm 0} = 0$ case because the SiO emitting layer is thinner. The overall shock layer is smaller because of the higher ionization fraction, and the layer fraction where SiO exists is thinner because of its destruction by C$^+$. Interestingly, a degeneracy is evident between the \{SiM, $G_{\rm 0} = 1$, 10\%\} and \{SiM, $G_{\rm 0} = 0$, 1\%\} results.
\end{itemize}

The most important conclusion from Fig.~\ref{figure1ant} is that it is possible to account for the observed levels of SiO emission in the frame of our assumptions. This figure shows that if the above constraint on the preshock density is accepted, the shock velocity is relatively well constrained, between 7 and 10\,km.s$^{-1}$ in the SiG scenario, and 10 and 12\,km.s$^{-1}$ in the SiM scenario. Such velocity ranges are well below the typical shock velocity required to sputter the grain cores ($\sim$25\,km.s$^{-1}$), allowing us to write off this SiO formation path. On the other hand, the radiation field, and the silicon distribution or chemistry is not well constrained by this comparison. However our results are encouraging and we will generalize our method to various lines of SiO and other molecules in order to constrain more input parameters. Combining the use of SiO with tracers of other physical quantities (such as the radiation field strength, which could be constrained through the observation of CO, C, and C$^+$ or H$_2$O, OH, and OI, or through the observations of ionized species) will help characterize the physical conditions within the observed shock regions.

\subsection{Shortcomings of our method}
\label{sec:soom}

From an observational point of view, the future work in this investigation will consist of the observation of additional SiO transitions to better constrain the models. A number of these transitions are available from ground-based facilities (1--0, 3--2, 5--4, 6--5, 8--7). Figure~\ref{figure4ant} shows that the SiO(2--1) transition is not necessarily optically thin in the models that are compatible with the observations, with a strong dependence on the considered model. In the future, lines with smaller opacities should be targeted, such as the (5--4), (6--5) and (8--7) lines. Of course, interferometric observations should be preferred in case distinct shock structures could be identified at higher angular resolution. In principle though, the knowledge of the filling factor of one transition is sufficient and can be generalized to the other transitions. We insist that SiO is a crucial molecule for such low-velocity shock environments. We will investigate the modeling of such emission in a forthcoming publication and show in another publication that the use of CO observations (such as performed in \citealt{gusdorf12,gusdorf15} in other Galactic shock environments) is much more complex. 

\begin{figure*}
\centering
\includegraphics[width=0.8\textwidth]{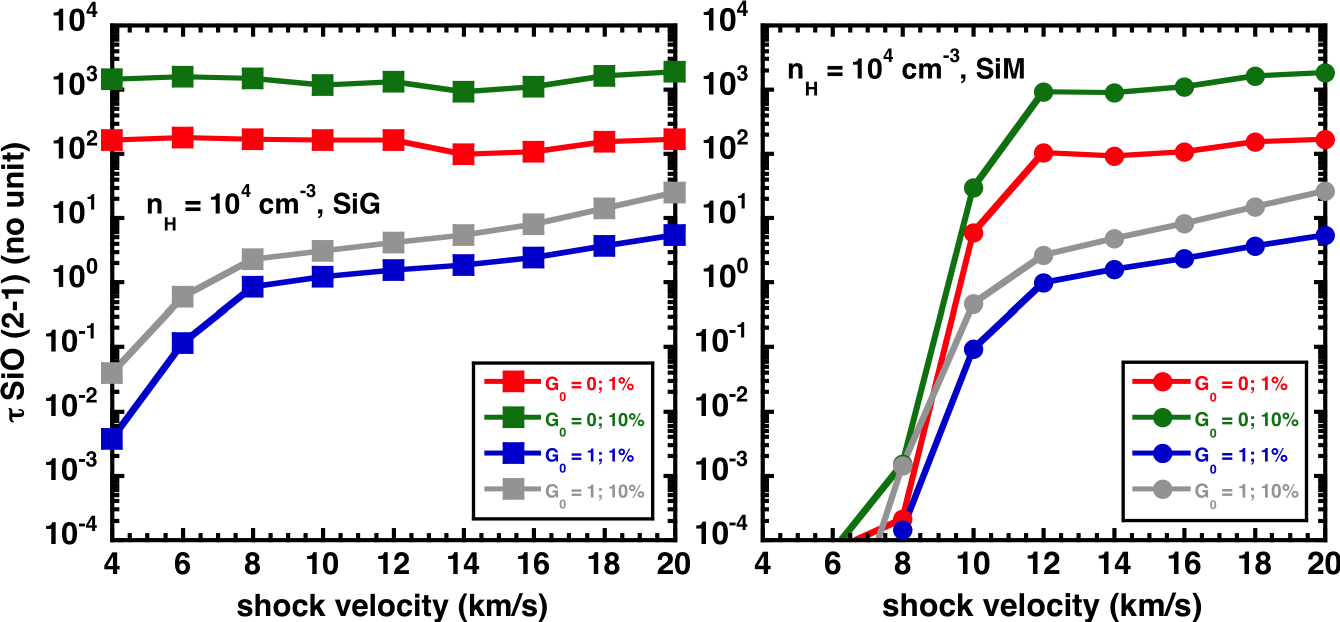}
\caption{ SiO(2--1) optical depth reached in each of the shock models, shown against the shock velocity, and calculated over the $n_{\rm H} = 10^4$~cm$^{-3}$ fragment of our grid (colored symbols). The left and right panels correspond to the SiG (square symbols) and SiM (circle symbols) scenarios (see text). The color code is indicated in the panels, where the $G_{\rm 0}$ (= 0 or 1) value and fraction of preshock free silicon (= 1 or 10 \%) is shown for each model.}
\label{figure4ant}
\end{figure*}

From the modeling point of view, it will be necessary enrich the chemical network. In particular, the development of photoreaction rate calculations will be crucial to our study, given the low velocity of the considered shocks. We have clearly displayed a region where the sputtering of the grains does not dominate all other processes as can be the case in more violent shock environments. The interaction of the UV radiation field with the gas, even at low $G_{\rm 0}$ value, can and will be investigated. In particular, the use of \textit{Herschel} observations of H$_2$O transitions will allow us to test the current understanding of the processes of photo- and thermal-desorption (e.g., \citealt{fayolle11a, fayolle11b, fayolle13, bertin13}) in this new shock environment.

\section{Conclusion}
\label{s:conclusion}

We mapped the high-mass star-forming ridge \object{W43-MM1} at high angular resolution with the IRAM Plateau de Bure Interferometer. We obtained SiO(2--1) and HCN(1--0) emission lines that we complemented with single-dish IRAM 30\,m radiotelescope data (see Fig.\ref{f:compa}). We used both transitions to look for outflows associated with star formation in \object{W43-MM1}. Our main results and conclusions may be summarized as follows:
\begin{itemize}
\item[$\bullet$] At high angular resolution, when observed with the IRAM PdBI interferometer, the SiO(2-1) emission appears very extended: $\sim$5\,pc in projected distance.
\item[$\bullet$] We retrieve evidence of seven outflows that were all detected both in SiO(2-1) and HCN(1-0) line transitions. Six of these outflows are associated with massive dense cores previously identified by \cite{louvet14}, pointing to the protostellar nature of these objects. The other outflow is associated with a lower mass object, the source \emph{G} identified by \citet{sridharan14}.
\item[$\bullet$] From the line profile analysis of the SiO(2-1) emission on the entire filament, we attribute half of the emission to high-velocity shocks linked with the stellar formation. The second half of the line emission appears to be very homogeneous along the \object{W43-MM1} filament with a mean FWHM of $\sim$6\,km.s$^{-1}$.
\item[$\bullet$] We characterize the line emission profile by defining eight independent areas along the filament and discover that four of the positions are only composed of one narrow component line profile (width of $\sim$6\,km.s$^{-1}$), one is composed of one single broad component line profile (source N12 that displays an outflow), and three are composed of both broad\, and \,narrow component line profiles. In the latter case, the narrow component is generally similar to those observed at the narrow-only positions, suggesting a common origin for this narrow component.
\item[$\bullet$] We run dedicated Paris-Durham shock models to confront the narrow component of the SiO(2-1) emission profile with SiO emission from low-velocity shocks. From this analysis, we show that the SiO emission intensity observed for the narrow component can be reproduced with shocks at low velocities in the range 7\,km.s$^{-1}$ to 12\,km.s$^{-1}$. These model-constrained velocities are in perfect agreement with the observational constraints of the SiO line width, which suggest velocities of shock between 4\,km.s$^{-1}$ to  14\,km.s$^{-1}$.
\end{itemize}

\section*{Acknowledgments}

This paper is based on observations carried out with the IRAM Plateau de Bure Interferometer. IRAM is supported by INSU/CNRS (France), MPG (Germany) and IGN (Spain).


\bibliographystyle{aa}
\bibliography{fab-these}

\end{document}